\documentclass[aps,prc,showpacs,superscriptaddress]{revtex4-1}
\usepackage{amsmath}
\usepackage{epsfig}
\usepackage{color}
\usepackage{endnotes}
\usepackage{natbib}
\usepackage[colorlinks,breaklinks]{hyperref}
\usepackage{nicefrac}

\usepackage{bm}
\usepackage{dcolumn}
\usepackage{graphics}
\usepackage{graphicx} 
\usepackage{tikz} 
\usepackage{amsmath}
\usepackage{amssymb} 
\usepackage{color}
\usepackage{changes}
\usepackage{multirow}
\usepackage{lineno}
\usepackage[normalem]{ulem}
\usepackage{xcolor}

\usepackage{sidecap}

\newcommand*{\TAU }{School of Physics and Astronomy, Tel Aviv University, Tel Aviv, Israel, 6997801}
\begin{document}

\title{Modeling Short-Range Nucleon Pair and Triplet Abundances in Atomic Nuclei}

\author{I. Wischnevsky Shlush}
\affiliation{\TAU}
\author{A. Denniston}
\affiliation{\TAU}
\author{R. Wagner}
\affiliation{\TAU}
\author{I. Korover}
\email[Contact Author \ ]{(korover@tauex.tau.ac.il)}
\affiliation{\TAU}
\author{E. Piasetzky}
\affiliation{\TAU}

\begin{abstract}
Short-range correlated (SRC) nucleon pairs provide a sensitive probe of the short-distance structure of atomic nuclei and the underlying nucleon-nucleon (NN) interaction. We present a simple numerical method to estimate the number of two- and three-nucleon SRC clusters using an independent-particle shell model with a harmonic-oscillator basis. The relative abundances of proton-neutron (pn), proton-proton (pp), and neutron-neutron (nn) SRC pairs are calculated for Al, Fe, and Pb nuclei, normalized to carbon, and compared with existing analytical predictions and available extractions from experimental data. We extend the analysis to the isotopes $^{40}$Ca, $^{48}$Ca, and $^{54}$Fe (collectively, the CaFe nuclei), which have been recently measured but whose SRC results are not yet published. These isotopes provide insight into the role of 1f7/2 shell occupancy in SRC pair formation. Finally, we present a reference baseline for three-nucleon SRC cluster abundances across the studied nuclei, assuming 3N clusters originate from independent two-nucleon interactions. The predicted baseline is about 2.5\% 3N-SRC/2N-SRC ratio for all medium and heavy nuclei. Deviations from this baseline may indicate the presence of additional short-range nuclear dynamics. 
\end{abstract}

\maketitle

Since the 1950s, the independent-particle shell model has served as a foundational framework for understanding nuclear structure~\cite{RevModPhys.77.427}. In this model, nucleons (protons and neutrons) move independently within distinct quantum orbitals (shells), each nucleon experiences an average potential generated by the remaining A-1 nucleons. While the nuclear shell model accurately predicts numerous nuclear properties it fails to fully account for experimental observations. Electron scattering experiments reveal that only 60–70\% of the expected number of protons occupy shell-model states~\cite{PhysRevC.54.2547,LAPIKAS1993297}. Modern calculations that include long-range correlations increase this fraction to approximately 80\%~\cite{DICKHOFF2004377}, but a significant deficit remains. This missing strength is attributed to short-range correlations (SRCs) nucleons, which involve strongly interacting nucleon pairs beyond the reach of mean-field approximations.
SRCs arise when two and three nucleons come into close spatial proximity such that the short-range components of the nucleon-nucleon (NN) interaction dominate over the long-range mean field~\cite{Frankfurt1988,ATTI20151}. These pairs are characterized by large relative momenta (above the fermi sea level, $k_F$) and small center-of-mass (CM) momenta, leading to depletion of shell-model occupancy and a redistribution of strength to high-momentum components—accounting for up to 20\% of nucleons in a nucleus.  Remarkably, SRCs appear to be universal across different nuclei. \\
Experimental studies using (e,e$^\prime$pN) reactions—where a high-momentum proton is knocked out and its partner is detected—show that nearly all high-momentum nucleons belong to SRC pairs, and $\sim 90\%$ of those are neutron-proton (np) pairs~\cite{subedi2008}. This is attributed to the dominance of the tensor component of the NN interaction at short distances, which favors np correlations over proton-proton (pp) or neutron-neutron (nn) pairs. 
Understanding SRCs is essential for several reasons:

\begin{itemize}
    \item \textbf{Beyond the Shell Model:} SRCs represent correlations beyond mean-field descriptions, thereby enriching our understanding of nuclear dynamics.
    \item \textbf{Astrophysical Relevance:} The high-density, short-range interactions in SRC pairs resemble conditions in neutron stars. Studying SRCs thus provides valuable input for modeling dense nuclear matter in astrophysical environments.
    \item \textbf{QCD Connections:} At very short distances, NN interactions probe regimes governed by quantum chromodynamics (QCD). SRC studies therefore offer an indirect way to access QCD effects in nuclear matter.
    \item \textbf{EMC Effect:} A potential link between SRCs and the EMC effect—the modification of parton distributions in bound nucleons - has been suggested~\cite{EMC2011}, indicating a deeper connection between nuclear structure and the quark-gluon substructure of nucleons.
\end{itemize}

Our objective here is to develop a simplified, predictive model for estimating the number of short-range correlated (SRC) nucleon pairs (2N-SRC) in atomic nuclei. The model is intentionally designed with a minimal set of adjustable parameters-only two in this case-enabling both a concise description of available experimental data and a predictive capability for forthcoming measurements.\\

\underline{\textbf{Model Overview}}\\
The first step in our model is to determine the probability ($P_2$) of finding two nucleons in spatial proximity, defined by a short-range correlation (SRC) volume characterized by a size parameter $r_{\mathrm{src}}$. The second key parameter is the probability for a SRC pair to be formed within this SRC volume, which accounts for the isospin-dependent nature of the nucleon-nucleon (NN) short range interaction. We assume:
\begin{itemize}
    \item $P_{np}(S=1) =1 $ i.e., a proton and a neutron with aligned spins always forms a correlated SRC pair when inside the SRC volume.
    \item $P_{nn}=P_{pp}=P_{np} (S=0)=p=0.22$ to reflect the lower pairing probability for spin=0 pairs (adjustable parameter).
\end{itemize}

\begin{figure}[ht]
    \centering
    \includegraphics[width=0.5\textwidth]{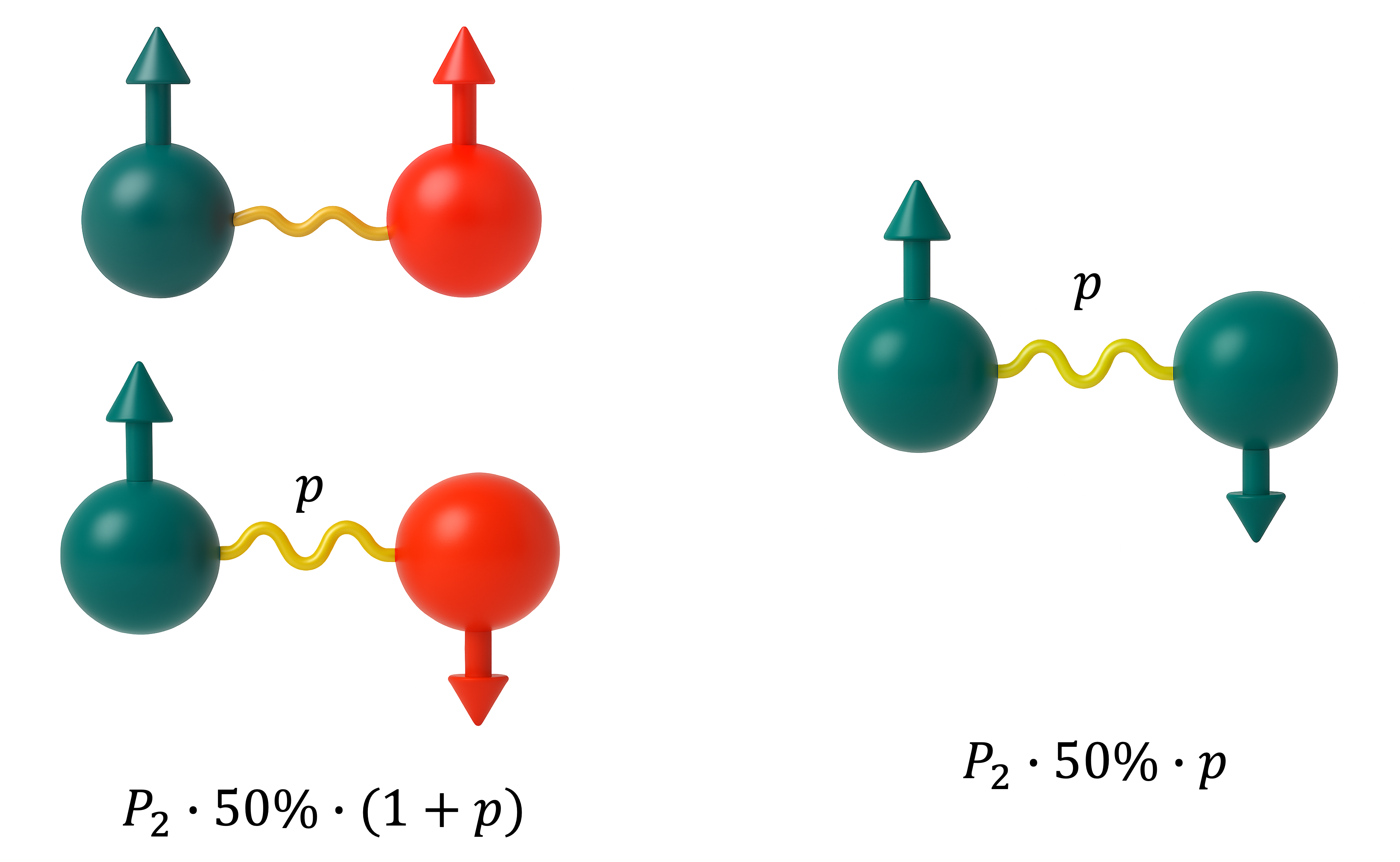}
    \caption{Calculating the probability of having a different isospin (left) and a same isospin SRC pair in a $r_{\mathrm{src}}$ cell. $P_2$ is the probability of having two nucleons in the cell and p the reduction for spin anti-parallel pairs. See text for details.}
    \label{fig:gen_discr}
\end{figure}

Assuming conditioned probability we calculate the probably $p2$ as shown in Fig.~\ref{fig:gen_discr}.\\

\textbf{Nuclear Wavefunction Model}\\
We model each nucleus using a three-dimensional harmonic oscillator potential, with radial wavefunctions of the form:
\begin{equation}
R_{n_r l}(r) = N_{n_r l} \left( \frac{r}{b} \right)^{l} 
L_{n_r}^{\left(l+\frac{1}{2}\right)}\!\left( \frac{r^2}{b^2} \right)
\exp\!\left( -\frac{r^2}{2b^2} \right)
\label{Eq:wave_func}
\end{equation}

where:\\[4pt]
\begin{tabular}{@{}ll@{}}
$n_r = 0, 1, 2, \ldots$ & is the radial quantum number,\\
$l$ & is the orbital angular momentum quantum number,\\
$L_{n_r}^{\alpha}(x)$ & is the associated Laguerre polynomial,\\
$b = \sqrt{\frac{\hbar}{m \omega}}$ &
\parbox[t]{0.72\textwidth}{is the oscillator length, where $m$ is the
proton mass and $\omega$ is the angular frequency of the oscillator,}\\
$N_{n_r l} = \sqrt{\frac{2\,n_r!}{b^3\,\Gamma(n_r + \frac{3}{2})}}$ &
is the normalization constant chosen such that.\\
\end{tabular}

\begin{equation}
\int_{0}^{\infty} \left| R_{n_r l}(r) \right|^{2} r^{2}\, dr = 1.
\end{equation}

\textbf{Shell Structure and Spatial Sampling}\\
The shell structure of each studied nucleus - $^{12}$C, $^{27}$Al, $^{40}$Ca, $^{48}$Ca, $^{54}$Fe,$^{56}$Fe, and $^{208}$Pb - was defined using standard shell-model configurations for protons and neutrons. 

\begin{table}[htbp]
\centering
\caption{Filled proton and neutron shells for selected nuclei.}
\renewcommand{\arraystretch}{1.2}
\setlength{\tabcolsep}{4pt}

\begin{tabular}{@{}l c l c l@{}}

\textbf{Nucleus} & \textbf{$Z$} &
\textbf{Proton Shells Filled} &
\textbf{$N$} &
\textbf{Neutron Shells Filled} \\

$^{12}\mathrm{C}$  & 6 &
\parbox[t]{0.28\textwidth}{1s$_{1/2}$(2), 1p$_{3/2}$(4)} &
6 &
\parbox[t]{0.28\textwidth}{1s$_{1/2}$(2), 1p$_{3/2}$(4)} \\[3pt]

$^{27}\mathrm{Al}$ & 13 &
\parbox[t]{0.28\textwidth}{1s$_{1/2}$(2), 1p$_{3/2}$(4), 1p$_{1/2}$(2), 1d$_{5/2}$(5)} &
14 &
\parbox[t]{0.28\textwidth}{1s$_{1/2}$(2), 1p$_{3/2}$(4), 1p$_{1/2}$(2), 1d$_{5/2}$(6)} \\[3pt]

$^{40}\mathrm{Ca}$ & 20 &
\parbox[t]{0.28\textwidth}{%
1s$_{1/2}$(2), 1p$_{3/2}$(4), 1p$_{1/2}$(2), 1d$_{5/2}$(6), 2s$_{1/2}$(2), 1d$_{3/2}$(4)} &
20 &
\parbox[t]{0.28\textwidth}{Same as protons} \\[3pt]

$^{48}\mathrm{Ca}$ & 20 &
\parbox[t]{0.28\textwidth}{Same as in $^{40}\mathrm{Ca}$} &
28 &
\parbox[t]{0.28\textwidth}{Adds 1f$_{7/2}$(8)} \\[3pt]

$^{54}\mathrm{Fe}$ & 26 &
\parbox[t]{0.28\textwidth}{Same as $^{40}\mathrm{Ca}$ + partial 1f$_{7/2}$(6)} &
28 &
\parbox[t]{0.28\textwidth}{1s$_{1/2}$ to 1f$_{7/2}$ fully filled} \\[3pt]

$^{56}\mathrm{Fe}$ & 26 &
\parbox[t]{0.28\textwidth}{Same as $^{54}\mathrm{Fe}$} &
30 &
\parbox[t]{0.28\textwidth}{Adds 2p$_{3/2}$(2)} \\[3pt]

$^{208}\mathrm{Pb}$ & 82 &
\parbox[t]{0.28\textwidth}{1s$_{1/2}$ to 3s$_{1/2}$ fully filled} &
126 &
\parbox[t]{0.28\textwidth}{1s$_{1/2}$ to 3p$_{1/2}$ fully filled} \\

\end{tabular}
\end{table}

We embed the entire nucleus within a cubic volume of side $\pm 10~\mathrm{fm}$, chosen to encompass the full radial distribution even for heavy nuclei such as $^{208}\mathrm{Pb}$ (with a nuclear radius estimated as 
$r \approx 1.24\,A^{1/3} \approx 7.4~\mathrm{fm}$).
This volume is discretized into smaller cubes with edge length $r_{\mathrm{src}}$, ensuring full coverage without leftover regions.
The centers of these sub-cubes serve as evaluation points for the nucleon wavefunction probabilities.

The oscillator length $b$ is adjusted independently for protons and neutrons using the empirical formulae and their average was used:
\begin{itemize}
    \item ${(\hbar \omega)}_{p} \approx 41 \cdot \left({Z} \right)^{-1/3}$\quad \text{for protons,}\\[6pt]
    \item $(\hbar \omega)_{n} \approx 41 \cdot \left({N} \right)^{-1/3}$\quad \text{for neutrons,}\\[6pt]
    \item $b = 0.5\left[(\sqrt{\frac{\hbar}{m \omega}})_{p} +(\sqrt{\frac{\hbar}{m \omega}})_{n}\right]$
\end{itemize}

To minimize parameter dependance and preserve a consistent physical interpretation across nuclei, we adopt a single averaged harmonic-oscillator length $b$ for both protons and neutrons. Although using separate $b_p$ and $b_n$ values provides a more formal description of the individual proton and neutron distributions, we find that the average-$b$ approach yields better agreement with experimental neutron-skin measurements for $^{48}$Ca and $^{208}$Pb. Specifically, the calculated neutron-skin thicknesses obtained with a unified $b$ reproduce the measured values within their reported uncertainties, whereas type-dependent oscillator lengths lead to an overestimation of the effect. This improved consistency supports the interpretation of $b$ as an effective mean spatial scale governing both proton and neutron wave-functions in medium and heavy nuclei, thereby justifying its use throughout this work.\\

\underline{\textbf{Calculation of SRC pair production}}\\
For each cell of size $r_{\mathrm{src}}$, we calculate the probability for it to be occupied by a single nucleon. 
As a consistency check, we sum the single-particle probabilities over all sub-cubes and verify that they reproduce 
the correct total numbers of protons and neutrons for each nucleus.

Next, we calculate the joint probability for two nucleons to reside within the same SRC volume. 
This allows extraction of probabilities for forming specific types of SRC pairs: $np$, $pp$, and $nn$, 
using the parameters presented above. 
We assume that $50\%$ of the second nucleons have the same spin orientation as the first nucleon in the cell 
and $50\%$ have opposite spin directions.

Finally, we scan over various combinations of the parameters 
$r_{\mathrm{src}} = 0.9$ - $1.1~\mathrm{fm}$ and $p=0.1$ - $0.35$ 
to identify configurations where:
\begin{enumerate}
  \item the total number of SRC pairs per nucleon, $\#\mathrm{pairs}/A$, 
  falls within the range of $15$--$20\%$, and
  \item the $pp/np$ ratio lies within $5$--$10\%$.
\end{enumerate}

In Refs.~\cite{RYCKEBUSCH19961,Colle2014,Colle2015}, the electron scattering cross sections in kinematics probing 
short-range correlated pairs were factorized into three components: 
(i) the cross section for virtual-photon absorption on a correlated SRC pair 
(depending on the relative coordinate of the two nucleons in the pair), 
(ii) a distorted two-body center-of-mass momentum distribution of the correlated SRC pair, 
and (iii) final-state interactions (FSIs) calculated in the RMSGA approximation. 
The factorized cross section was evaluated in Ref.~\cite{Colle2015} within the zero-range approximation (ZRA), 
which corresponds to setting the relative pair coordinate to zero, 
thereby selecting the very short-range components of the relative SRC pair wave functions.

The absolute number of SRC pairs is not an observable. 
However, the \emph{ratio} of SRC pairs between two nuclei can be extracted from 
hard electron-scattering measurements performed in SRC-sensitive kinematics. 
Inclusive, semi-exclusive, and exclusive cross-section ratios, 
together with model-dependent corrections, can be used to infer SRC-pair ratios. 
A full extraction of SRC-pair numbers from data is beyond the scope of this work; 
instead, we compare our SRC-pair numbers with values reported in the literature.

\begin{figure}
    \centering
    \includegraphics[width=0.8\linewidth]{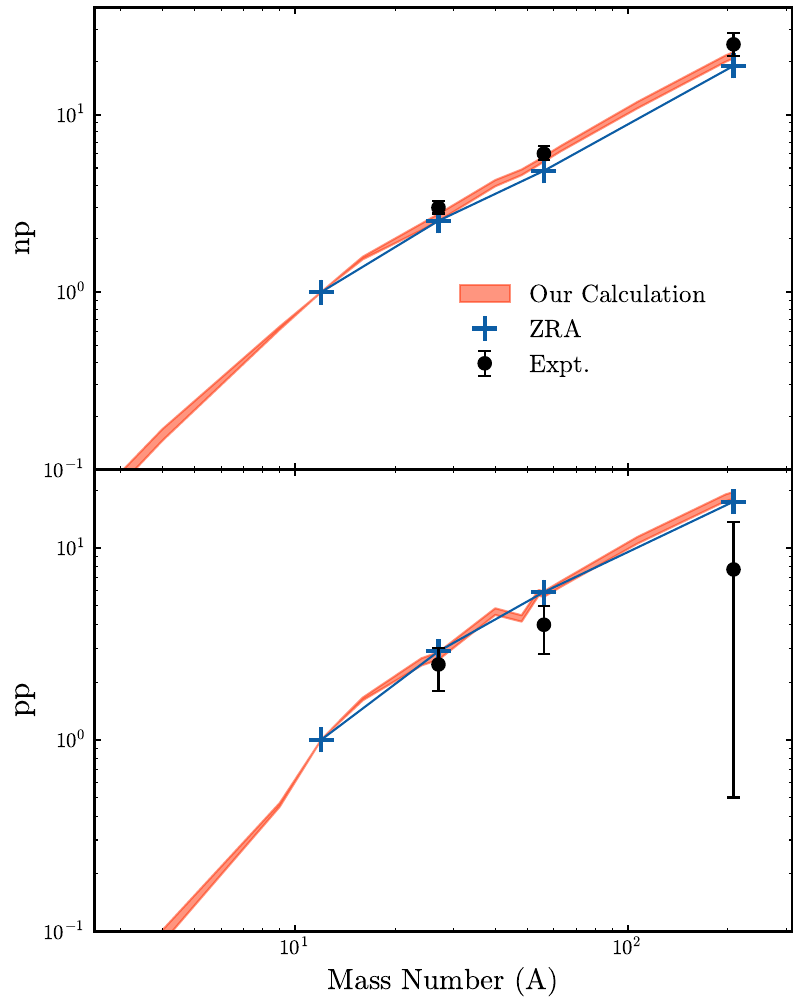}
    \caption{Mass (A) dependence of the number of np (top panel) and pp (bottom panel) SRC pairs of nucleus A relative to $^{12}$C.  Data (small full squares with error bars) are as extracted in ~\cite{Colle2015} from the measured CLAS A(e,e$^\prime$p) and A(e,e$^\prime$pp) cross-section ratios~\cite{EMC2011,Fomin2012} after correcting for FSI. Error bars include the estimated uncertainty on the cross-section ratios and the FSI corrections. The empty crosses denote the result of the ZRA calculation of ref~\cite{Colle2015} Our model calculations in the relevant ranges of the adjustable parameters ($p$, $r_{\mathrm{src}}$) are presented as a band (red online).}
    \label{fig:pair_abund}
\end{figure}

In Fig.~\ref{fig:pair_abund}, we compare the ZRA calculation of Ref.~\cite{Colle2015} and the relative pair abundances 
extracted from experimental data~\cite{Colle2015} with our simulated results obtained for 
$r_{\mathrm{src}} = 0.9$ - $1.1~\mathrm{fm}$ 
and assuming equal probabilities for $nn$ and $pp$ pairs, $p = 0.1$--$0.35$.

\begin{figure}
    \centering
    \includegraphics[width=0.8\linewidth]{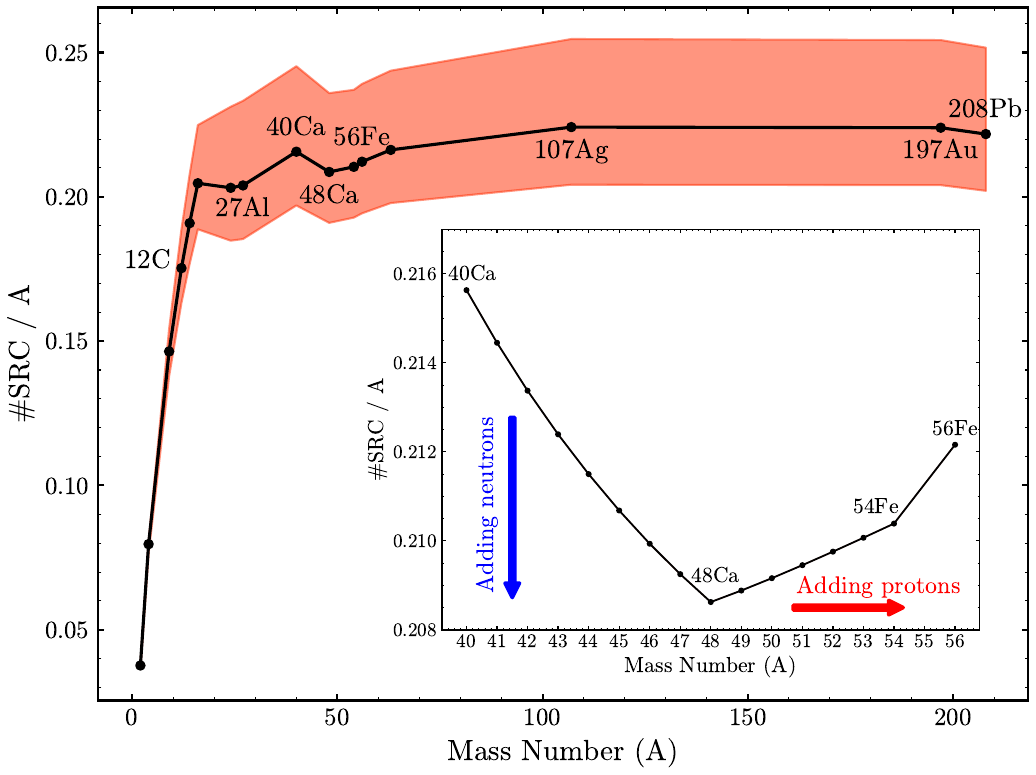}
    \caption{Number of SRC pairs as calculated by us as a function of A. The band (red online) represents the range of adjustable parameter as discussed above. The nominal value (black line) is calculated using the nominal parameters $r_{\mathrm{src}}=1\ \mathrm{fm}$ and $p=0.22$, which are obtained by requiring (pp/np)=$7.5\%$ and $\#$pairs/A=$17.5\%$. }
    \label{fig:All_nucl}
\end{figure}

With the model parameters as defined above, in Fig.~\ref{fig:All_nucl} we present our calculated number of SRC pairs per nucleon over the periodic table from deuteron to lead. 

In the model, as the atomic number A increase so does the harmonic oscillator parameter $b$ (see eq. above). As $b$ increase so the size of the potential wall and the space spread of the wave functions that represent the protons and neutrons in the nucleus. That cause a reduction of the probability of two nucleons to get close together and yield the overall SRC/A curve which gets asymptotically flat at large A. Superimpose on the global A dependence observed in the figure the opening of new shells (f shell for $A=40$) is shown in more details in the insert. Adding more neutrons to the new external shells with no proton in them do not add many np- SRC pairs and is observed as local minima in the SRC/A ratio. See for examples $^{40}$Ca vs. $^{48}$Ca Adding protons to these shells increase the number of SRC pairs. See for example $^{54}$Fe and $^{56}$Fe compared to $^{48}$Ca.\\
Recently, two electron-scattering experiments at Jefferson Lab measured the relative number of high-momentum protons and neutrons in the nuclei of $^{40}$Ca, $^{48}$Ca, and $^{54}$Fe - collectively referred to as the "CaFe nuclei". By comparing $^{40}$Ca and $^{48}$Ca, one can isolate the effect of adding eight neutrons in the 1f7/2 shell on the probability of forming short-range correlated (SRC) nucleon pairs. Similarly, the transition from $^{48}$Ca to $^{54}$Fe, which involves adding six protons to the same external 1f7/2 shell, allows us to examine SRC pairing between the added protons and both the core nucleons and the 1f7/2 neutrons. Therefore, comparing the extracted number of SRC pairs in the CaFe nuclei offers insight into SRC pairing within and across nuclear shells.\\
Using our model, we predict the number of SRC pairs in these nuclei under the assumption that all nucleons - whether in the core or in the external 1f7/2 shell - contribute to SRC pair formation if they get close to each other (share the same $r_{\mathrm{src}}$ box). This serves as a baseline for comparison with experimental results. Any deviations may point to either enhanced or suppressed pairing involving external-shell nucleons. The model calculations, using the adjustable parameters discussed earlier, are summarized in Table~\ref{tab:src_pairs}.

\begin{table}[htbp]
\centering
\caption{Model-predicted number of SRC pairs in the Ca and Fe nuclei.}
\renewcommand{\arraystretch}{1.2} 
\setlength{\tabcolsep}{10pt}      
\begin{tabular}{|l|ccc|}
\hline
\textbf{Nucleus} & \textbf{$np$ pairs} & \textbf{$pp$ pairs} & \textbf{$nn$ pairs} \\ 
\hline\hline
$^{40}\mathrm{Ca}$ & 7.37 & 0.629 & 0.629 \\
$^{48}\mathrm{Ca}$ & 8.49 & 0.580 & 0.940 \\
$^{54}\mathrm{Fe}$ & 9.68 & 0.791 & 0.887 \\ 
\hline

\end{tabular}
\label{tab:src_pairs}
\end{table}

As seen in Table~\ref{tab:src_pairs}, adding eight neutrons (a 40\% increase) to $^{40}$Ca to form $^{48}$Ca increases the number of nn-SRC pairs by approximately 50\%, the number of np-SRC pairs by about 15\%, and slightly reduced  the number of pp-SRC pairs. Subsequently, adding six protons to $^{48}$Ca - a 30\% increase in proton number - to obtain $^{54}$Fe increases the number of pp-SRC pairs by 35\%, the number of np-SRC pairs by 14\%, while the number of nn-SRC pairs  being reduced slightly.
These predictions reflect a baseline scenario in which nucleons in the 1f7/2 shell form SRC pairs with the $^{40}$Ca core nucleons as core-core nucleon pairs, based solely on the spatial overlap and short-range interaction strength. However, if selection rules or other mechanisms suppress pairing between external shell nucleons and core nucleons, the ratios of np- and nn- SRC pairs in $^{48}$Ca relative to $^{40}$Ca would be lower than predicted. Conversely, if the 1f7/2 shell nucleons are more likely to form SRC pairs, for example due to specific spin or angular momentum couplings, the np-SRC pair ratio in $^{54}$Fe relative to $^{48}$Ca could exceed our model prediction.
Within  the model calculation we can also itemize the SRC pairs according to the shells that the nucleons that compose them originated from. For $^{12}$C, $^{40}$Ca, $^{48}$Ca, and $^{54}$Fe that information is given in Tables~\ref{tab:onlyCarb},\ref{tab:HeavyNuc}. These pairs are produced based on having two nucleons close to each other without quantum number selection. Such selection can reduce or even eliminate contributions from specific shell combinations.\\

\begin{table}[ht]
    \centering
    \caption{Number of pairs per-shell for $^{12}$C.}
    \begin{tabular}{|l|ccc|}
    \hline
    \textbf{Type} & \textbf{np} & \textbf{pp} & \textbf{nn} \\
    \hline\hline
    ss & 0.258 & 0.012 & 0.012 \\
    sp & 0.853 & 0.077 & 0.077 \\
    pp & 0.718 & 0.049 & 0.049 \\
    \hline
    \end{tabular}
    \label{tab:onlyCarb}
\end{table}

\begin{table}[h!]
\centering
\caption{Number of pairs per-shell for $^{40}$Ca, $^{48}$Ca, and $^{54}$Fe.}
\begin{tabular}{|l|ccc|ccc|ccc|}
\hline
\textbf{Type} & \multicolumn{3}{c|}{$^{40}$Ca} & \multicolumn{3}{c|}{$^{48}$Ca} & \multicolumn{3}{c|}{$^{54}$Fe} \\
 & np & pp & nn & np & pp & nn & np & pp & nn \\
\hline\hline
ss & 0.421 & 0.027 & 0.027 & 0.401 & 0.026 & 0.026 & 0.386 & 0.025 & 0.025 \\
sp & 1.221 & 0.111 & 0.111 & 1.134 & 0.103 & 0.103 & 1.071 & 0.097 & 0.097 \\
pp & 0.897 & 0.068 & 0.068 & 0.821 & 0.062 & 0.062 & 0.768 & 0.058 & 0.058 \\
sd & 1.338 & 0.121 & 0.121 & 1.220 & 0.110 & 0.110 & 1.137 & 0.103 & 0.103 \\
pd & 2.054 & 0.186 & 0.186 & 1.883 & 0.170 & 0.170 & 1.766 & 0.160 & 0.160 \\
dd & 1.436 & 0.117 & 0.117 & 1.337 & 0.109 & 0.109 & 1.268 & 0.103 & 0.103 \\
sf &  &  &  & 0.309 &  & 0.056 & 0.502 & 0.039 & 0.052 \\
pf &  &  &  & 0.509 &  & 0.092 & 0.842 & 0.065 & 0.087 \\
df &  &  &  & 0.880 &  & 0.159 & 1.468 & 0.114 & 0.152 \\
ff &  &  &  &  &  & 0.053 & 0.474 & 0.027 & 0.050 \\
\hline
\end{tabular}
    \label{tab:HeavyNuc}
\end{table}

\begin{SCfigure}[1][!htbp]
\caption{Calculating the probability of having a npp or nnp SRC triplet in a $r_{\mathrm{src}}$ cell. $P_3$ is the probability of having three nucleons in the cell and p the reduction for spin anti-parallel pairs. See text for details.}
\includegraphics[width=0.28\textwidth]{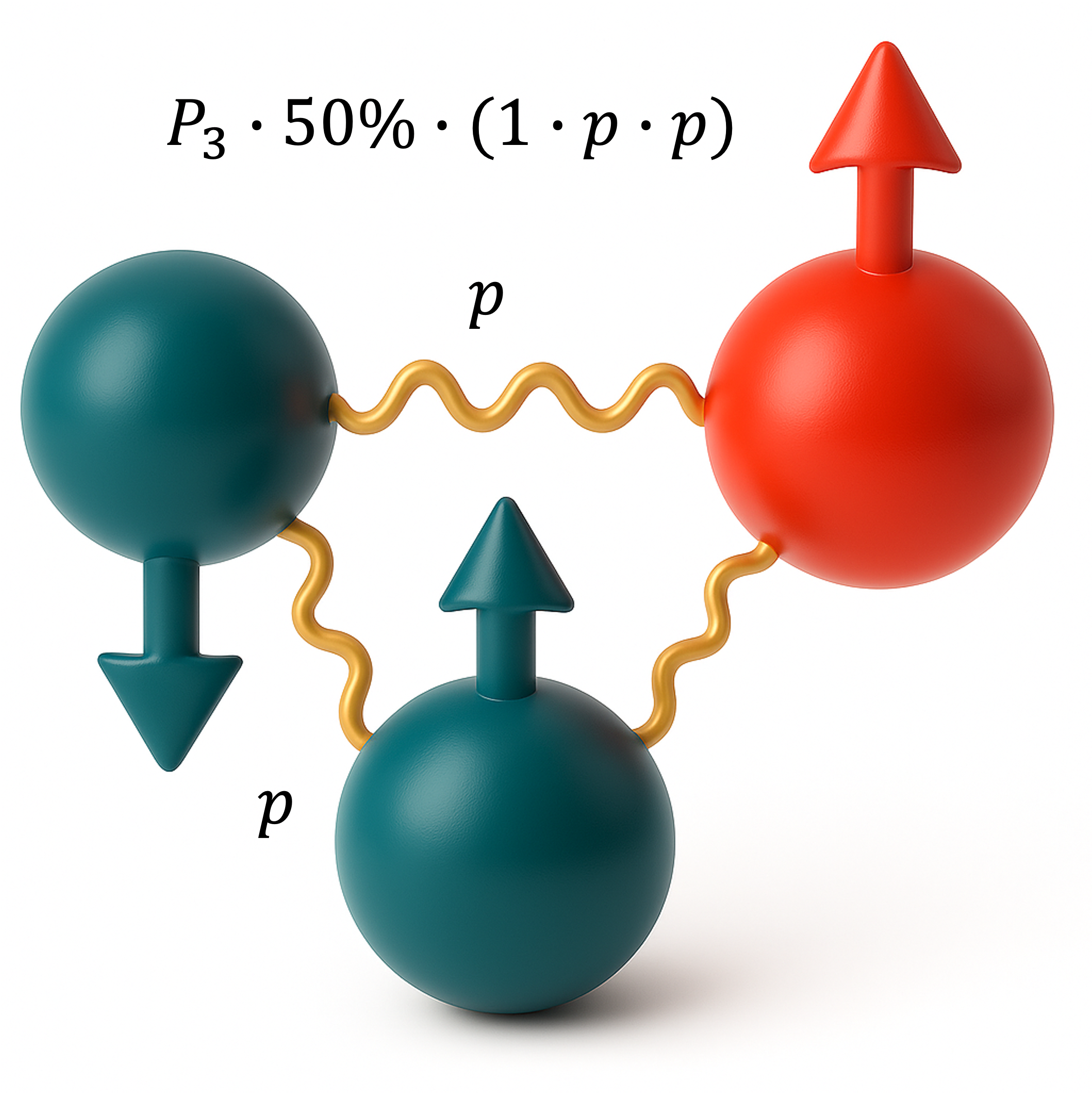}
 \label{fig:3N_conf}
\end{SCfigure}

\underline{\textbf{Three nucleons SRC triplets}}\\
Unlike 2N-SRC, the features and importance of 3N-SRC are mostly unknown. Given the probability of forming a 3N-SRC is significantly lower, a direct measurement of 3N-SRC is a challenge that requires enormous luminosity, beam time, and ability to identify the 3N-SRC signal on a large background of other processes. The electron scattering measurements~\cite{Nguyen2020,Egiyan2006,Fomin2012,Ye3Nsearch,DDay2023} yield so far inconclusive results. Identifying and measuring the 3N-SRC correlation is  the leading challenge of the field now.\\
Following similar approach to count 2N-SRC correlation we predict here the number of different isospin structure SRC triplet in nuclei. This predication can serve as a baseline to study the process that creates SRC triplet clusters from the mean filed nucleons and study its similarity and differences from the SRC pairs production. \textbf{Our reference baseline assumes that the 3N SRC triplets are due to 2N forces only and that the force between any two nucleons is independent of other nuclei in the nucleus}. 
The first step in our model is to determine the probability ($P_3$) of finding three nucleons in spatial proximity, defined by a short-range correlation (SRC) volume characterized by a radius parameter $r_{\mathrm{src}}$. We used the independent shell model wave functions presented above for the 2N SRC calculations. The second key parameter is the probability for a SRC triplet to be formed within this SRC volume, which accounts for the isospin-dependent nature of the nucleon-nucleon (NN) short range interaction. Fig.~\ref{fig:3N_conf} shows the assumed probability for 3 nucleons to create 3N-SRC triplet. We assume that p is as defined above for 2N-SRC. In this model we summed no \textbf{ppp-SRC} and \textbf{nnn-SRC} triplet which require same isospin nucleons with parallel spins.

\begin{figure}[!htbp]
    \centering
    \includegraphics[width=0.72\linewidth]{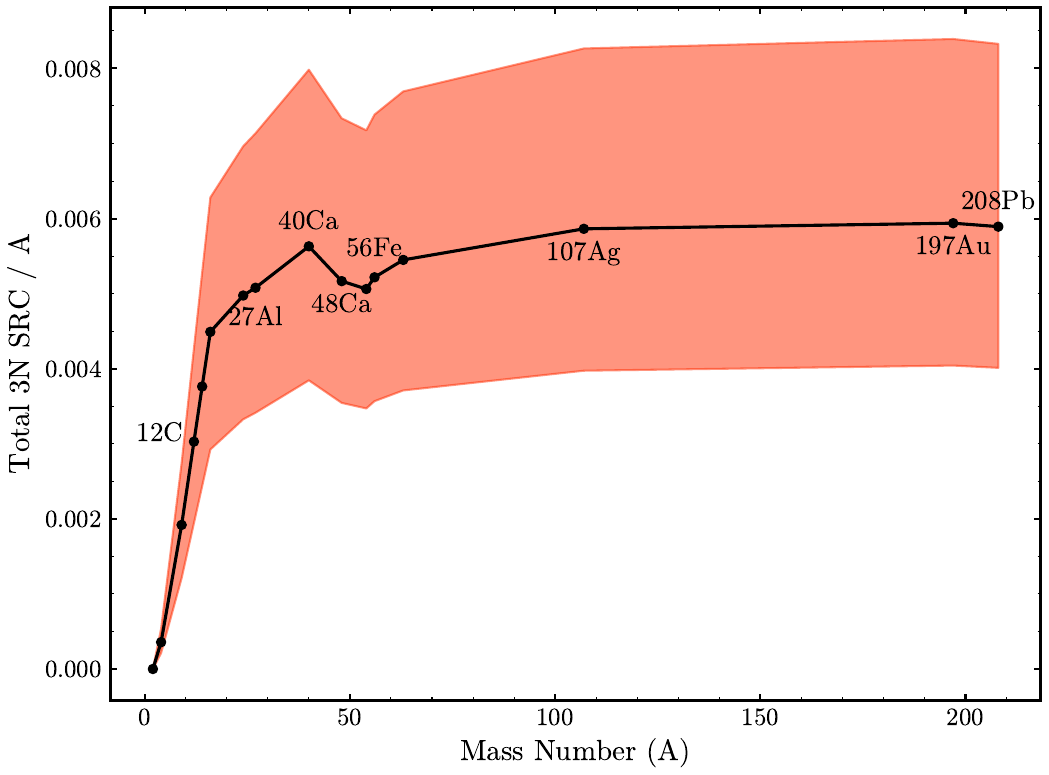}
    \caption{The predicted 3N-SRC/A ratio as a function of the mass number. The calculations assumes that the 3N-SRC triplets are due to 2N forces only and that the force between any two nucleons is independent of other nucleons in the nucleus.See text for details.}
    \label{fig:3N_results}
\end{figure}

As can be seen in Fig.~\ref{fig:3N_results} the baseline assumption yield for all medium and heavy nuclei a constant 3N/2N ratio of about 2.5\%. This extends the universality to 3N SRC and is expected assuming that the local short distance is independent of the nucleus. Note that authors of Ref. [17] did not explicitly calculate the 3N-SRC/2N-SRC ratio. However, they
did evaluate the mass dependence of the 3N-SRC contribution (see Fig. 5 in Ref.~\cite{Ryckebusch2012}), finding it to increase monotonically with $A^{1.5}$. Since the 2N-SRC strength scales roughly as $A$, their results imply that the 3N-SRC/2N-SRC ratio should increase with the mass number in contrast to our model.\\

\underline{\textbf{Summary and Conclusions}}\\
In this work, we presented a simplified yet effective model to estimate the number of short-range correlated (SRC) nucleon pairs based on shell-model ground state configurations. Despite its simplicity, the model reproduces key trends observed in experimental data and provides predictive power for new measurements. Its success lies in leveraging the spatial overlap of shell-model nucleon wavefunctions while incorporating a minimal number of parameters tuned to capture isospin-dependent pairing probabilities.
The calculated relative abundances of SRC pairs in nuclei such as Al, Fe, and Pb were shown to be consistent with both values extracted from experimental data and previous zero-range approximation (ZRA) calculations~\cite{Colle2015}. \\
We extended our predictions to the CaFe nuclei which have recently been studied experimentally but whose SRC data are not yet available. Our predictions provide a baseline under the assumption that nucleons in the external 1f7/2 shell contribute to SRC pair formation in the same way as core nucleons do. If external shell nucleons participate in SRC formation differently than core nucleons, this will manifest as measurable deviations from the predicted np pp and nn SRC pair ratios, and correspondingly, in the (e,e$^\prime$), (e,e$^\prime$p), and (e,e$^\prime$n) cross-section ratios.
We further used our SRC pair estimates to calculate 3N SRC triplets. These predictions can be compared with upcoming experimental results to isolate potential deviations in SRC behavior linked to shell-specific effects.

\begin{acknowledgments}
The authors are grateful for fruitful discussions with Lawrence Weinstein, Or Hen, and Julian Kahbow.  This research was supported by the Israeli Science Foundation (ISF) under grants No. 917/20(E.P.), No. 371/23  (E.P), No. 830/24 (I. K.), the ISF NSFC joint research program grant No. 3107/23 (E. P.), and the PAZY Foundation 520/23 (E.P.), No. 737/25 (I.K.).
\end{acknowledgments}

\bibliographystyle{apsrev4-1}

\bibliography{references.bib}

\end{document}